\documentclass[12pt]{article}
\usepackage{graphicx}
\usepackage{lscape}
\usepackage{citesort}

\begin{document}

\def\ds{\displaystyle}
\def\beq{\begin{equation}}
\def\eeq{\end{equation}}
\def\bea{\begin{eqnarray}}
\def\eea{\end{eqnarray}}
\def\beeq{\begin{eqnarray}}
\def\eeeq{\end{eqnarray}}
\def\ve{\vert}
\def\vel{\left|}
\def\ver{\right|}
\def\nnb{\nonumber}
\def\ga{\left(}
\def\dr{\right)}
\def\aga{\left\{}
\def\adr{\right\}}
\def\lla{\left<}
\def\rra{\right>}
\def\rar{\rightarrow}
\def\nnb{\nonumber}
\def\la{\langle}
\def\ra{\rangle}
\def\ba{\begin{array}}
\def\ea{\end{array}}
\def\tr{\mbox{Tr}}
\def\ssp{{\Sigma^{*+}}}
\def\sso{{\Sigma^{*0}}}
\def\ssm{{\Sigma^{*-}}}
\def\xis0{{\Xi^{*0}}}
\def\xism{{\Xi^{*-}}}
\def\qs{\la \bar s s \ra}
\def\qu{\la \bar u u \ra}
\def\qd{\la \bar d d \ra}
\def\qq{\la \bar q q \ra}
\def\gGgG{\la g^2 G^2 \ra}
\def\q{\gamma_5 \not\!q}
\def\x{\gamma_5 \not\!x}
\def\g5{\gamma_5}
\def\sb{S_Q^{cf}}
\def\sd{S_d^{be}}
\def\su{S_u^{ad}}
\def\ss{S_s^{??}}
\def\sbp{{S}_Q^{'cf}}
\def\sdp{{S}_d^{'be}}
\def\sup{{S}_u^{'ad}}
\def\ssp{{S}_s^{'??}}
\def\sig{\sigma_{\mu \nu} \gamma_5 p^\mu q^\nu}
\def\fo{f_0(\frac{s_0}{M^2})}
\def\ffi{f_1(\frac{s_0}{M^2})}
\def\fii{f_2(\frac{s_0}{M^2})}
\def\O{{\cal O}}
\def\sl{{\Sigma^0 \Lambda}}
\def\es{\!\!\! &=& \!\!\!}
\def\ap{\!\!\! &\approx& \!\!\!}
\def\ar{&+& \!\!\!}
\def\ek{&-& \!\!\!}
\def\kek{\!\!\!&-& \!\!\!}
\def\cp{&\times& \!\!\!}
\def\se{\!\!\! &\simeq& \!\!\!}
\def\eqv{&\equiv& \!\!\!}
\def\kpm{&\pm& \!\!\!}
\def\kmp{&\mp& \!\!\!}


\def\simlt{\stackrel{<}{{}_\sim}}
\def\simgt{\stackrel{>}{{}_\sim}}


\title{
         {\Large
                 {\bf
Analysis of rare $B \rar K_0^\ast (1430) \ell^+ \ell^-$ decay
within QCD sum rules
                 }
         }
      }

\author{\vspace{1cm}\\
{\small T. M. Aliev \thanks
{e-mail: taliev@metu.edu.tr}~\footnote{permanent address:Institute
of Physics,Baku,Azerbaijan}\,\,,
K. Azizi\,\,,
M. Savc{\i} \thanks
{e-mail: savci@metu.edu.tr}} \\
{\small Physics Department, Middle East Technical University,
06531 Ankara, Turkey} }

\date{}

\begin{titlepage}
\maketitle
\thispagestyle{empty}

\begin{abstract}
Form factors of rare $B \rar K_0^\ast (1430) \ell^+ \ell^-$ decay are
calculated within three--point QCD sum rules, with $K_0^\ast (1430)$ being
the p--wave scalar meson. The branching ratios are estimated when only
short, as well as short and long distance effects, are taken into account.
It is obtained that the $B \rar K_0^\ast (1430) \ell^+ \ell^-$ $(\ell =
e,\mu)$ decay is measurable at LHC.
Measurement of these branching ratios for the semileptonic rare 
$B \rar K_0^\ast (1430) \ell^+ \ell^-$ can give valuable information about
the nature of scalar meson $K_0^\ast (1430)$.
\end{abstract}

\end{titlepage}

\section{Introduction}

The flavor changing neutral current (FCNC) processes induced by $b \rar
s(d)$ transitions provide the most sensitive and stringiest test for the
standard model (SM) at one loop level, since they are forbidden in SM at
tree level \cite{R8401,R8402}. For this reason these decays are very
sensitive to the physics beyond the SM via the influence of new particles in
the loops.

Despite the branching ratios of FCNC decays are small in the SM, quite
intriguing results are obtained in ongoing experiments. The inclusive
$B \rar X_s \ell^+ \ell^-$ decay is observed in BaBaR \cite{R8403} and Belle
collaborations. These collaborations also measured exclusive
modes $B \rar K \ell^+ \ell^-$ \cite{R8404,R8405,R8406} and 
$B \rar K^\ast \ell^+ \ell^-$ \cite{R8407}. The experimental results on
these decays are in a good agreement with theoretical estimations
\cite{R8408,R8409,R8410}.

There is another class of rare decays induced by $b\rar s$ transition,
such as $B \rar K_{02}^\ast (1430) \ell^+ \ell^-$ in which B meson decays
into p--wave scalar meson. The decays $B \rar K_2^\ast (1430) \ell^+
\ell^-$ and $B \rar K_0^\ast (1430) \ell^+ \ell^-$ are studied in
\cite{R8411} and \cite{R8412}, respectively. The main problem in these
studies is the calculation of the transition form factors. Transition form
factors of these decays in the framework of light front quark
model \cite{R8412} are estimated in \cite{R8413} and \cite{R8414}, 
respectively.

In the present work we calculate the transition form factors of $B \rar
K_0^\ast (1430) \ell^+ \ell^-$ decay in 3--point QCD sum rules method
\cite{R8415} (for a review on the QCD sum rules method, see \cite{R8416}),
where $K_0^\ast (1430)$ is the p--wave scalar meson. The main reason for
studying $B \rar K_0^\ast (1430) \ell^+ \ell^-$ decay form factors in the
framework of 3--point QCD sum rules method is that part of the form factors,
namely, those appearing in calculation of the matrix element of the axial
vector current between initial and final meson states, can easily be
obtained from the results for the $B_s \rar D_s (2317)$ transition form 
factors \cite{R8421}, which are calculated within the same
framework, with the help of appropriate replacements (see section 2). 
The quark structure of the scalar mesons have not
been unambiguously determined yet and is still under discussion. In order to
establish the inner structure of scalar mesons much more effort from
theoretical and experimental sides are needed.
This method is employed to calculate the form factors of heavy meson
semileptonic decays, $D \rar K^0 e \bar{\nu}_e$ \cite{R8417},
$B \rar D(D^\ast) \ell \bar{\nu}_\ell$ \cite{R8418},
$D \rar K(K^\ast) e \bar{\nu}_e$ \cite{R8419}, $D \rar \pi e
\bar{\nu}_e$ \cite{R8420}, $B \rar K(K^\ast) \ell^+ \ell^-$ (first reference
in \cite{R8408}), etc. 

The work is organized as follows: In section 2, we calculate the relevant
hadronic form factors of the $B \rar K_0^\ast (1430) \ell^+ \ell^-$ decay
with the help of 3--point QCD sum rules. Section 3 is devoted to the numerical 
analysis and discussion of the considered decay and our conclusions.

\section{Form factors of the $B \rar K_0^\ast (1430) \ell^+ \ell^-$ decay}

The exclusive $B \rar K_0^\ast (1430) \ell^+ \ell^-$ decay is described at
quark level by $b \rar s \ell^+ \ell^-$ transition. The effective
Hamiltonian responsible for the $b \rar s \ell^+ \ell^-$ transition in the
standard model is
\bea
\label{e8401}
{\cal H}_{eff} \es {G_F \alpha V_{tb} V_{ts}^\ast \over 2\sqrt{2} \pi} \Bigg[
C_9^{eff} (m_b) \bar{s}\gamma_\mu (1-\gamma_5) b \, \bar{\ell} \gamma^\mu \ell
+ C_{10} (m_b) \bar{s} \gamma_\mu (1-\gamma_5) b \, \bar{\ell} \gamma^\mu
\gamma_5 \ell \nnb \\
\ek 2 m_b C_7 (m_b) {1\over q^2} \bar{s} i \sigma_{\mu\nu}
(1+\gamma_5) b \, \bar{\ell} \gamma^\mu \ell \Bigg]~,
\eea
where $C_7$, $C_9^{eff}$ and $C_{10}$ are the Wilson coefficients.
It is well known that the Wilson coefficient $C_9^{eff}$ has a perturbative
part as well as a resonant part which comes from the long--distance effects 
due to the conversion of the real $\bar{c}c$ into lepton pair (see, for 
example the second reference in \cite{R8408}) which can be written as
\bea
C_9^{eff} = C_9^{per} + C_9^{res}~. \nnb
\eea
The explicit expressions of $C_7$, $C_9^{per}$ and $C_{10}$ can be found in
\cite{R8401}. $C_9^{res}$ is usually parametrized by using Breit--Wigner
ansatz,
\bea
C_9^{res} = {3\pi \over \alpha^2} C^{(0)} \sum_{V_i=\psi(1s) \cdots
\psi(6s)} \ae_i \, {\Gamma(V_i \rar \ell^+ \ell^-) m_{V_i} \over m_{V_i}^2-
q^2-im_{V_i}\Gamma_{V_i}} ~, \nnb
\eea
where $\alpha$ is the fine structure constant and $C^{(0)}=0.362$.    

The phenomenological factors $\ae_i$ for the $B \rar K(K^\ast) \ell^+ \ell^-$
decay can be determined from the condition that they should reproduce
correct branching ratio relation
\bea
{\cal B} (B \rar J/\psi K(K^\ast) \rar K(K^\ast) \ell^+ \ell^-) = 
{\cal B} (B \rar J/\psi K(K^\ast)) {\cal B} (J/\psi \rar \ell^+ \ell^-)~, \nnb
\eea
where the right--hand side is determined from experiments. Using the
experimental values of the branching ratios for the $B \rar V_i K(K^*)$ and
$V_i \rar \ell^+ \ell^-$ decays, for the lowest two $J/\psi$ and $\psi^\prime$
resonances, the factor $\ae$ takes the values: $\ae_1=2.7,~\ae_2=3.51$ (for
$K$ meson), and $\ae_1=1.65,~\ae_2=2.36$ (for $K^\ast$ meson). The values of
$\ae_i$ used for higher resonances are usually the average of the values
obtained for the $J/\psi$ and $\psi^\prime$ resonances.  
But, unfortunately
the mode $B \rar J/\psi K_0^\ast(1430)$ for the $B \rar K_0^\ast(1430) \ell^+
\ell^-$ decay has not been seen yet and therefore the right--hand side of
the above--mentioned equation for the  $B \rar K_0^\ast(1430) \ell^+
\ell^-$ decay is unknown. For this reason we are not able to determine the
factors $\ae_i$ for the $B \rar K_0^\ast(1430) \ell^+ \ell^-$ decay.

As we have already mentioned above, the values of the factors $\ae_i$ are of
order $\ae\sim 1$ or $\ae\sim 2$, for the $B \rar K \ell^+ \ell^-$ and $B \rar
K^\ast \ell^+ \ell^-$ decays. For this reason, even though qualitatively, in order 
to determine the branching ratio for the $B \rar K^\ast_0(1430) \ell^+ \ell^-$ 
decay with the inclusion of long distance effects, the values of $\ae_i$ we
use are borrowed from $B \rar K \ell^+ \ell^-$ and $B \rar
K^\ast \ell^+ \ell^-$ decays, namely, we choose $\ae=1$ and $\ae=2$ and
performed numerical calculations with these values. 

The amplitude for the 
$B \rar K_0^\ast (1430) \ell^+ \ell^-$ decay can be obtained after evaluating
the matrix elements of the quark operators in Eq. (\ref{e8401}) between
initial $B$ and final $K_0^\ast (1430)$ meson states. It follows from Eq.
(\ref{e8401}) that in order to obtain the amplitude for the $B \rar K_0^\ast
(1430) \ell^+ \ell^-$ decay the matrix elements $\lla K_0^\ast  \vel
\bar{s}\gamma_\mu (1-\gamma_5) \ver B \rra$ and $\lla K_0^\ast  \vel
\bar{s} i \sigma_{\mu\nu} q^\mu (1+\gamma_5)  \ver B \rra$ are needed. 
These matrix elements are parametrized in terms of the form factors as
follows:
\bea
\label{e8402}
\lla K_0^\ast (1430) (p^\prime) \vel \bar{s}\gamma_\mu \gamma_5 b \ver 
B(p) \rra \es f_+ (q^2) {\cal P}_\mu + f_-(q^2) q_\mu~, \\
\label{e8403}
\lla K_0^\ast (1430) (p^\prime) \vel \bar{s}i \sigma_{\mu\nu} q^\nu
\gamma_5 b  \ver B(p) \rra \es 
{f_T (q^2) \over m_B + m_{K_0^\ast}} \big[ {\cal P}_\mu q^2 - 
(m_B^2 - m_{K_0^\ast}^2 ) q_\mu \big]~,
\eea
where ${\cal P}_\mu = (p+p^\prime)_\mu$ and $q_\mu = (p-p^\prime)_\mu$.
For calculation of these form factors in the framework of the QCD sum rules
method, we consider the following three--point correlators
\bea   
\label{e8404}
\Pi_\mu(p,p^\prime,q) \es - \int d^4x \, d^4y e^{i(p^\prime y - px)}
\lla 0 \vel {\rm T} J^{K_0^\ast}(y) J_\mu^A(0) J_5^B(x) \ver 0 \rra~, \\
\label{e8405}
\Pi_\mu(p,p^\prime,q) \es - \int d^4x \, d^4y e^{i(p^\prime y - px)}
\lla 0 \vel {\rm T} J^{K_0^\ast}(y) J_{\mu\nu}(0) J_5^B(x) \ver 0 \rra~,
\eea
where $J^{K_0^\ast}=\bar{d}s$,  $J_5^B=\bar{b} i\gamma_5 d$ and
$J_{\mu\nu}=\bar{s} i \sigma_{\mu\nu} b$ are the interpolating currents of
the scalar $K_0^\ast (1430)$, and $B$ mesons and weak flavor changing quark
currents, respectively. The expressions of the form factors $f_+(q^2)$ and
$f_-(q^2)$ can be obtained from the results for the $B_s \rar D_{s_0}
(2317)$ transition form factors given in \cite{R8421}, with the
help of the replacements $f_{D_{s_0}} \rar f_{K_0^\ast}$, $m_{D_{s_0}} \rar
m_{K_0^\ast}$, $m_c \rar m_s$, and $m_s \rar m_d=0$. For completeness we
also present the sum rules for the form factors $f_+(q^2)$ and $f_-(q^2)$
below. 

For this reason we proceed by describing in full detail the
calculation of $F_T$ by considering the correlator given in Eq.
(\ref{e8405}). This correlator can be decomposed into a set of the following
independent Lorentz structures:
\bea
\label{e8406}
\Pi_{\mu\nu} = (p_\mu p_\nu^\prime - p_\nu p_\mu^\prime) \Pi_T +
\sum_n a_{\mu\nu}^{(n)} \Pi^{(n)}~,
\eea
where $\Pi$ and $\Pi^{(n)}$ are functions of $p^2$, $p^{\prime 2}$ and
$q^2$, and $a_{\mu\nu}^{(n)}$ are the other tensors which can be built using
the vectors $p$ and $p^\prime$ and the metric tensor $g_{\mu\nu}$.

In further analysis we will be interested only on the invariant amplitude
$\Pi_T$, for whose amplitude we write the dispersion relation
\bea
\label{e8407}
\Pi_T(p^2,p^{\prime 2},q^2) = - \frac{1}{(2 \pi)^2} \int_{m_b^2}^{\infty} ds 
\int_{m_s^2}^{\infty} ds^\prime
\frac{\rho (s,s^\prime,Q^2)}{(s-p^2) (s^\prime-p^{\prime 2})} +
\mbox{\rm subtraction terms}~,
\eea
where $\rho (s,s^\prime,Q^2)$ is the spectral density, $Q^2=-q^2$.
According to the main idea of the QCD sum rules method, left hand side of
Eq. (\ref{e8407}) must be calculated at large Euclidean momenta $p^2$ and 
$p^{\prime 2}$ with the help of operator product expansion (OPE). The right
hand side of Eq. (\ref{e8407}) is determined by saturating it with the
lowest mesonic states. Applying then the Borel transformation on the
variables $p^2$ and $p^{\prime 2}$ suppresses the higher resonance
contributions and higher power corrections, as a result of which we get the
sum rules for the corresponding quantities.

Let us start by calculating the hadronic part of the correlator
(\ref{e8405}). Saturating this correlator with $B$ and $K_0^\ast(1430)$
mesons and selecting the structure $(p_\mu p_\nu^\prime - p_\nu
p_\mu^\prime)$, we get
\bea
\label{e8408}
\Pi_T = - {2 f_{K_0^\ast} m _{K_0^\ast} \over m_B + m _{K_0^\ast}} {f_B
m_B^2 \over m_b} {f_T(q^2) \over (m _{K_0^\ast}^2 - p^{\prime 2})
(m_B^2-p^2)}~.
\eea
In deriving Eq. (\ref{e8408}) we use
\bea
\label{e8409}
\lla 0 \vel J^{K_0^\ast} \ver K_0^\ast \rra \es  f_{K_0^\ast} m_{K_0^\ast}~, \\
\label{e8410}
\lla B \vel J_5^B \ver 0 \rra \es {f_B m_B^2 \over m_b}~,\\
\label{e8411}
\lla K_0^\ast \vel \bar{s} i \sigma_{\mu\nu} b \ver B \rra \es  - 2 (p_\mu
p_\nu^\prime - p_\nu p_\mu^\prime) {f_T(q^2) \over m_B + m _{K_0^\ast}}~.
\eea

The theoretical part of the correlator (\ref{e8405}) can be calculated with
the help of OPE at short distance. Up to operators having dimension $6$, it
is determined by the bare loop and power corrections from operators with
$d=3~\lla \bar{q}q \rra$, $d=4~\lla G^2 \rra$, $d=5~m_0^2 \lla \bar{q}q \rra$
and $d=6~\lla \bar{q}q \rra^2$. After calculating all power corrections, we
obtain that $\lla G^2 \rra$ and $\lla \bar{q}q \rra^2$  give negligibly
small correction and for this reason we do not present their explicit
expressions.

The spectral density for bare loop in Eq. (\ref{e8407}) can be obtained by
replacing the denominator $1/(p^2-m^2)$ of the quark propagators by $-2\pi i
\delta(p^2-m^2)$ in the initial Feynman integral. After standard calculation
we get
\bea
\label{e8412}
\rho_T (p^2,p^{\prime 2},Q^2) = {N_c \over \lambda^{3/2} } 
\Big\{ m_b \ga 2 s^\prime \Delta - \Delta^\prime u \dr - 
m_s \ga 2 s \Delta^\prime - \Delta u \dr \Big\}~,
\eea
where $N_c=3$ is the color factor, $\Delta=s-m_b^2$, $\Delta^\prime =
s^\prime - m_s^2$, $u=s+s^\prime+Q^2$, $\lambda(p^2,p^{\prime 2},Q^2) = u^2
- 4 s s^\prime$.

The integration region for the perturbative contribution in Eq.
(\ref{e8407}) is determined from the condition that the argument of all
three $\delta$ functions might vanish simultaneously. the bounds of $s$ and
$s^\prime$ are determined from the following inequality:
\bea
\label{e8413}
-1 \le {u \Delta - 2 s \Delta^\prime \over 
\lambda^{1/2}(s,s^\prime,Q^2) \Delta} \le +1~.
\eea
The contributions of higher states is parametrized as the corresponding
spectral density starting from $s>s_0$ and $s^\prime>s_0^\prime$ invoking
quark hadron duality. 

We now present the results of calculations of dimension--3 and --5
operators:
\bea
\label{e8414}
\Pi_T^{(3)} \es - {\lla \bar{q}q \rra \over r r^\prime}~, \\
\label{e8415}
\Pi_T^{(5)} \es - m_0^2 \lla \bar{q}q \rra \Bigg[
{1 \over 3 r^2 r^\prime} + {1 \over 3 r^{\prime 2} r } - {m_b^2 \over 2 r^3
r^\prime} - {m_s^2 \over 2 r  r^{\prime 3} }
- {2 m_b^2 + 2 m_s^2 - 2 m_b m_s + 2 Q^2 \over 12 r^2 r^{\prime 2}
}\Bigg]~,
\eea 
where $r = p^2 - m_b^2$, $r^\prime = p^{\prime 2} - m_s^2$.

In our calculations we do not calculate ${\cal O}(\alpha_S)$ corrections to
the bare loop, and for consistency we also neglect ${\cal O}(\alpha_S)$
corrections in leptonic decay constants $f_{K_0^\ast}$ and $f_B$.

Substituting Eqs. (\ref{e8408}), (\ref{e8414}) and (\ref{e8415}) into Eq.
(\ref{e8407}) and performing double Borel transformation in the variables
$p^2$ and $p^{\prime 2}$, we get the following sum rule for the form factor
$f_T$:
\bea
\label{e8416} 
f_T \es - { m_B + m_{K_0^\ast} \over 2 f_{K_0^\ast} m_{K_0^\ast} }
{m_b \over f_B m_B^2} e^{(m_B^2/M_1^2 + m_{K_0^\ast}^2/M_2^2 )}        
\Bigg\{ - {1 \over (2 \pi)^2 } \int ds \, ds^\prime \rho_T (s,s^\prime,Q^2)
e^{-( s/M_1^2 + s^\prime/M_2^2 )} \nnb \\
\ar e^{-(m_b^2/M_1^2 + m_s^2/M_2^2 )}
\Bigg[ - \lla \bar{q}q \rra + m_0^2 \lla \bar{q}q \rra \Bigg( 
{1 \over M_1^2} + {1 \over 3 M_2^2} + {m_b^2 \over 4 M_1^4} + {m_s^2 \over 4
M_2^4} \nnb \\ 
\ar {2 m_b^2 - 2 m_b m_s + 2 m_s^2 + 2 Q^2 \over 12 M_1^2 M_2^2 } \Bigg)
\Bigg]\Bigg\}~,
\eea

For completeness we also present sum rules for the form factors $f_+$ and
$f_-$.
\bea
\label{e8417}
f_+ \es - {m_b \over f_B m_B^2} {1 \over f_{K_0^\ast} m_{K_0^\ast} }
e^{m_B^2/M_1^2 + m_{K_0^\ast}^2/M_2^2}
\Bigg\{
- {1 \over (2\pi)^2 } \int ds\, ds^\prime \rho_+ e^{-( s/M_1^2 +
s^\prime/M_2^2 )} \nnb \\
\ar e^{-(m_b^2/M_1^2 + m_s^2/M_2^2 )}
\Bigg[ {1\over 2} \lla \bar{q}q \rra (m_b-m_s)  -
{1\over 12} m_0^2 \lla \bar{q}q \rra \Bigg( { 3 m_b^2 (m_b-m_s) \over 2
M_1^4} +
{ 3 m_s^2 (m_b-m_s) \over 2  
M_2^4}\nnb \\ 
\ek {2 (m_b-2m_s) \over M_2^2} - {2 (2m_b-m_s) \over M_1^2} +
{(m_b-m_s) (2 m_b^2 + m_b m_s + 2 m_s^2 + 2 Q^2) \over M_1^2 M_2^2 }
\Bigg) \Bigg] \Bigg\}~, \\ \nnb \\
\label{e8418}
f_- \es - {m_b \over f_B m_B^2} {1 \over f_{K_0^\ast} m_{K_0^\ast} }
e^{m_B^2/M_1^2 + m_{K_0^\ast}^2/M_2^2}
\Bigg\{
- {1 \over (2\pi)^2 } \int ds\, ds^\prime \rho_- e^{-( s/M_1^2 +
s^\prime/M_2^2 )} \nnb \\
\ar e^{-(m_b^2/M_1^2 + m_s^2/M_2^2 )}
\Bigg[ - {1\over 2} \lla \bar{q}q \rra (m_b+m_s)  +
{1\over 12} m_0^2 \lla \bar{q}q \rra \Bigg( { 3 m_b^2 (m_b+m_s) \over 2
M_1^4} \nnb \\
\ar { 3 m_s^2 (m_b+m_s) \over 2
M_2^4} 
- {2 (m_b+3m_s) \over M_2^2} - {2 (3m_b+m_s) \over M_1^2} \nnb \\
\ar {(m_b+m_s) (2 m_b^2 + m_b m_s + 2 m_s^2 + 2 Q^2) \over M_1^2 M_2^2 }
\Bigg) \Bigg] \Bigg\}~,
\eea
where
\bea
\label{e8419}
\rho_+ \es {N_c \over4 \lambda^{1/2}}
\Big\{ (\Delta^\prime + \Delta) (1+A+B) +
[(m_b+m_s)^2 + Q^2 ] (A+B) \Big\}~ \\ \nnb \\
\label{e8420}
\rho_- \es {N_c \over4 \lambda^{1/2}}
\Big\{ [\Delta^\prime + \Delta +
(m_b+m_s)^2 + Q^2] (A-B) + \Delta^\prime - \Delta \Big\}~,
\eea
and
\bea
\label{e8421}
A \es {1 \over \lambda} [ 2 s^\prime \Delta
-u \Delta^\prime ]~, \nnb \\
B \es {1 \over \lambda} [ 2 s \Delta^\prime - u \Delta]~,
\eea
Using Eqs. (\ref{e8401})--(\ref{e8403}), we get the
following expression for the differential decay width:
\bea
\label{e8422}
{d \Gamma (B \rar K_0^\ast \ell^+ \ell^-) \over d\hat{s}} \es
{G^2 \alpha^2 m_B^5 \over 3072 \pi^5} \vel V_{tb} V_{ts}^\ast \ver^2
v \sqrt{\lambda(1,\hat{m}_{K_0^\ast}^2,\hat{s})} \Bigg\{ \Bigg[
\vel C_9^{eff} f_+ + {2 \hat{m}_b \over 1 + 
\hat{m}_{K_0^\ast}} C_7 f_T \ver^2 \nnb \\
\ar \vel C_{10} f_+ \ver^2 \Bigg] (3-v^2)
\lambda(1,\hat{m}_{K_0^\ast}^2,\hat{s}) +
12 \hat{m}_\ell^2 \Big[ (2 + 2 \hat{m}_{K_0^\ast}^2 -\hat{s}) 
\vel f_+ \ver^2 \nnb \\ 
\ar 2 (1-\hat{m}_{K_0^\ast}^2) {\rm Re} [f_+ f_-^\ast] + 
\hat{s} \vel f_- \ver^2 \vel \Big] C_{10} \ver^2
\Bigg\}~,
\eea
where 
\bea
\hat{s} = {q^2 \over m_B^2}~,~~~~~ v = \sqrt{1 - {4 \hat{m}_\ell^2 \over
\hat{s}}}~,~~~~~ \hat{m}_b = {m_b \over m_B}~,~~~~~
\hat{m}_\ell = {m_\ell \over m_B}~,~~~~~
\hat{m}_{K_0^\ast} = {m_{K_0^\ast} \over m_B}~, \nnb
\eea
and $\lambda(1,m_{K_0^\ast}^2,\hat{s}) = 1 + m_{K_0^\ast}^4 + \hat{s}^2 - 2
\hat{s} - 2 m_{K_0^\ast}^2 (1+\hat{s})$.

\section{Numerical results of the sum rules}

In this section we present our numerical study for the form factors
$f_+(q^2)$, $f_-(q^2)$, and $f_T(q^2)$ for the $B \rar K_0^\ast(1430) \ell^+
\ell^-$ decay. From the expressions of the sum rules for the form factors we
see that the main input parameters of all form factors are quark
condensates, leptonic decay constants $f_B$ and $f_{K_0^\ast}$ of $B$ and
$K_0^\ast$ mesons, respectively, Borel parameters $M_1^2$ and $M_2^2$, as
well as continuum thresholds $s_0$ and $s_0^\prime$.

The value of the quark condensate at $\mu=1~GeV$ is $\lla \bar{q}q \rra =
-(240 \pm 10~MeV)^3$, $m_c(\mu=m_c) = (1.275 \pm 0.015)~GeV$,
$m_b=(4.7\pm0.1)~GeV$ \cite{R8422}. The leptonic decay
constant of $B$ and the continuum thresholds $s_0$ and
$s_0^\prime$ are determined from the analysis of
two--point sum rules which have the values: $f_B=180~MeV$ \cite{R8416},
$s_0 = (35\pm 2)~GeV^2$ and $s_0^\prime = (4.4\pm 0.4)~GeV^2$ \cite{R8423}.
Few words about the value of the leptonic decay constant $f_{K_0^\ast}$ are in
order. This constant is calculated within the 2--point QCD sum rules method
in \cite{R8423}, including ${\cal O}(\alpha_s)$ corrections. As has already
been noted, we neglect ${\cal O}(\alpha_s)$ corrections in the bare loop
calculations, and hence, for consistency, we shall neglect them in
calculation of the leptonic decay constant $f_{K^\ast}$. Neglecting ${\cal
O}(\alpha_s)$ corrections from the result of \cite{R8423}, we obtain for the
leptonic decay constant $f_{K^\ast}=340~MeV$, which we shall use in further 
numerical calculations.

The Borel parameters $M_1^2$ and $M_2^2$ are not physical quantities. 
The results for the
physically measurable quantities should be independent of them if the OPE
can be performed up to infinite order. But, in sum rules, OPE is truncated
to some finite order and for this reason the Borel parameters have to be
chosen in such a working region that the physical results are practically
independent on them. In choosing the working region of $M_1^2$ and $M_2^2$
the conditions must be satisfied: 1) the contribution of the excited should
be small, and, 2) the power corrections should converge.

Our numerical analysis shows that both conditions are satisfied and the best
stabilities of all form factors are achieved when $M_1^2$ and $M_2^2$ vary
in the regions $8~GeV^2 \le M_1^2 \le 15~GeV^2$ and $2.5~GeV^2 \le M_2^2 \le
4.5~GeV^2$, respectively.

The values of the form factors at $q^2=0$ are
\bea
\label{e8423}
f_+(0) \es \phantom{-} 0.31 \pm 0.08~, \nnb \\
f_-(0) \es -0.31 \pm 0.07~, \nnb \\
f_T(0) \es -0.26 \pm 0.07~,
\eea
where the errors are due to the variation of Borel parameters, the continuum
thresholds $s_0$ and $s_0^\prime$, the uncertainty in the condensate
parameters, the variation of quark mass and meson decay constants. 

Here we would like to make the following cautionary note. From the sum rules
expressions for the form factors it is seen that they
are very sensitive to the magnitude of the leptonic decay constant
$f_{K_0^\ast}$ of the $K_0^\ast(1430)$ meson. The above--mentioned results
for $f_i(0)$ are obtained at $f_{K_0^\ast}=340~MeV$. But, if we use
$f_{K_0^\ast}=300~MeV$ which is given in \cite{R8414} (after normalizing it
to our definition) we obtain
\bea
\label{e8424}
f_+(0) \es \phantom{-} 0.34 \pm 0.08~, \nnb \\
f_-(0) \es -0.34 \pm 0.07~, \nnb \\
f_T(0) \es -0.29 \pm 0.07~.
\eea
When compared with our results, $f_+(0)$ and $f_T(0)$ are in good agreement
within the error limits, while the value of $f_-(0)$ is $50\%$ 
larger compared to the prediction of \cite{R8414}.
   
In calculating the total width of the $B \rar K_0^\ast(1430) \ell^+
\ell^-$ decay, we need to know the $q^2$ dependence of the form factors 
$f_+(q^2)$, $f_-(q^2)$ and $f_T(q^2)$ in the physical region $4 m_\ell^2 \le
q^2 \le (m_B-m_{K_0^\ast})^2$. The $q^2$ dependence of the form factors in
the physical region can be calculated directly from sum rules, which has
comprehensively been discussed in \cite{R8419,R8420} and \cite{R8424}. 
We are restricted to
consider a region of $q^2$, calculated from the QCD side, where the 
correlator can be reliable. Our numerical analysis shows that this region is
bounded as $0 \le q^2 \le 8~GeV^2$. In order to extend the results to full
physical region, we look such a parametrization of the form factors that
they coincide with the sum rule predictions in the above--mentioned
region of $q^2$. 

Our analysis shows that the best fit for the $q^2$ dependence of the form 
factors can be written in the following form:
\bea
\label{e8425}
f_i(\hat{s}) = {f_i(0)\over 1 - a_i \hat{s} + b_i \hat{s}^2}~,
\eea
where $i=+$, $-$ or $T$ and $\hat{s} = q^2/m_B^2$.
The values of the parameters $f_i(0)$, $a_i$ and $b_i$   are 
given in table 1.   
\begin{table}[h]    
\renewcommand{\arraystretch}{1.5}
\addtolength{\arraycolsep}{3pt}
$$
\begin{array}{|l|ccc|}  
\hline
& f_i(0) & a_i & b_i \\ \hline
f_+ &
\phantom{-}0.31 \pm 0.08 & 0.81 & -0.21 \\
f_- &
-0.31\pm 0.07 & 0.80 & -0.36 \\
f_T &
-0.26\pm 0.07 & 0.41 & - 0.32 \\ \hline
\end{array}
$$
\caption{Form factors for $B \rar K_0^\ast(1430) \ell^+
\ell^-$ decay in a three--parameter fit for $f_{K_0^\ast}=340~MeV$.}
\renewcommand{\arraystretch}{1}
\addtolength{\arraycolsep}{-3pt}
\end{table}
      
Using the equation of motion the form factor $f_T$ can be related to $f_-$
as follows:
\bea
\label{e8426}
f_T(q^2) = {(m_b-m_s) \over (m_B - m_{K^\ast})} \, f_- ~.
\eea
We see that, within the errors Eq. (\ref{e8426}) is in agreement with the
computed form factor $f_T$ from QCD sum rules.

The kinematical interval of the dilepton invariant mass $q^2$ is $4 m_\ell^2
\le q^2 \le (m_B - m_{K_0^\ast})^2$ in which the long distance effects (the
charmonium resonances) can give substantial contribution. The dominant
contribution to the $B \rar K_0^\ast (1430) \ell^+ \ell^-$ decays 
comes from the two low lying resonances $J/\psi$
and $\psi^\prime$, in the interval of $8~GeV^2\le
q^2 \le 14~GeV^2$. In order to minimize the hadronic uncertainties we
discard this subinterval by dividing the kinematical region of 
$q^2$ as follows:

\bea
\begin{array}{cl}
\mbox{\rm I} & 4 m_\ell^2 \le q^2 \le (m_{J\psi} - 0.02~GeV)^2~,\\ \\
\mbox{\rm II} & (m_{J\psi} + 0.02~GeV)^2 \le q^2 \le 
(m_{\psi^\prime} - 0.02~GeV)^2~, \\ \\
\mbox{\rm III} & (m_{\psi^\prime} + 0.02~GeV)^2 \le q^2 \le 
(m_B-m_{K_0^\ast})^2~. \nnb
\end{array} \nnb
\eea

In Fig. (1) we present the dependence of the differential branching ratio 
for the $B \rar K_0^\ast (1430) e^+ e^-$ decay on
$q^2$, as well as its dependence on $q^2$ due solely to short distance
effects ($\ae=0$ case).
 
Taking into account the $q^2$ dependence of the form factors given in Eq.
(\ref{e8425}), performing integration over $\hat{s}$, and using the total
life time $\tau_B = 1.53 \times 10^{-12}~s$ \cite{R8425}, we get the
following for the branching ratios when only short distance contribution 
is taken into account:
\bea
{\cal B} (B \rar K_0^\ast(1430) e^+ e^-) \es 2.09 \times 10^{-7}~(2.68
\times 10^{-7})~, \nnb \\
{\cal B} (B \rar K_0^\ast(1430) \mu^+ \mu^-) \es 2.07 \times 10^{-7}~(2.66
\times 10^{-7})~, \nnb \\
{\cal B} (B \rar K_0^\ast(1430) \tau^+ \tau^-) \es 1.70 \times 10^{-9}~(2.20
\times 10^{-9})~, \nnb
\eea
where the values in the parenthesis correspond to the choice of
$f_{K_0^\ast}=300~MeV$.

Taking long distance effects into account in the above--mentioned 
kinematical regions, we get the following for the branching ratios:

\bea
{\cal B}(B \rar K_0^\ast(1430) e^+ e^-) = \left\{ \begin{array}{lll}
1.97 \times 10^{-7}& \mbox{\rm region I}~,& \\ \\
1.66 \times 10^{-8}& \mbox{\rm region II}~,& \mbox{\rm at}~\ae=1,\\ \\ 
3.90 \times 10^{-10}&  \mbox{\rm region III}~,& \end{array} \right. \nnb
\eea
and
\bea
{\cal B}(B \rar K_0^\ast(1430) e^+ e^-) = \left\{ \begin{array}{lll}
2.13 \times 10^{-7}& \mbox{\rm region I}~,& \\ \\
1.63 \times 10^{-8}& \mbox{\rm region II}~,& \mbox{\rm at}~\ae=2,\\ \\
3.38 \times 10^{-10}&  \mbox{\rm region III}~,& \end{array} \right. \nnb
\eea
at $f_{K_0^\ast} = 340~MeV$. Note that the branching ratios of the $B \rar
K_0^\ast(1430) e^+ e^-$ and $B \rar K_0^\ast(1430) \mu^+ \mu^-$ decays are
practically the same.

It follows from these results that the dominant contribution comes from region I 
(low invariant mass region), and this can be attributed to the existence of the
factor $1/q^2$. Since at LHC--b $10^{11}$--$10^{12}$ pairs are
hoped to be produced, the expected number of events for the $B \rar
K_0^\ast(1430) e^+ e^-$ decay in the low invariant mass region
is of the order of $10^4$--$10^5$. Therefore, this mode sounds to be quite
measurable at LHC--b. Comparing the value of the branching ratio for the $B
\rar K_0^\ast(1430) e^+ e^-$ decay for the case when only short distance
effects are taken into account, with the value calculated in region I, it
can be said that they practically coincide. Hence, in the light of above
results, we can conclude that long distance effects do not give significant 
contribution in region I, and therefore, measurement of branching ratio in
this region allows us to check the short distance structure of the effective
Hamiltonian.
      
The smallness of the value of ${\cal B} (B \rar K_0^\ast(1430) \tau^+
\tau^-)$ can be attributed to the small phase volume of this decay. Taking
into account the efficiency for detecting $\tau$ leptons, the measurement of
the branching ratio for $B \rar K_0^\ast(1430) \tau^+ \tau^-$ is very
difficult, even at LHC. 

As the concluding remark we can state that, we study the semileptonic rare
$B \rar K_0^\ast(1430) \ell^+ \ell^-$ decay. The transition form factors of
the $B \rar K_0^\ast(1430) \ell^+ \ell^-$ are calculated in the framework of
three--point QCD sum rules. The branching ratios of the relevant decay for
$\ell=e, \mu, \tau$ leptons, when short and long distance effects are taken
into account, are estimated. From these results we conclude
that the $B \rar K_0^\ast(1430) \ell^+ \ell^-$ ($\ell=e,\mu$)
decay can be measured in future planned experiments at LHC.

\section*{Acknowledgments} 

One of the authors (T. M. A) is grateful to T\"{U}B\.{I}TAK for partially
support of this work under the project 105T131.

\newpage

\section*{Figure captions}
{\bf Fig. (1)} The dependence of the differential branching ratio
for the $B \rar K_0^\ast (1430) e^+ e^-$ decay on
$q^2$, for the values of the fudge factor $\ae=0$ (corresponding to short
distance effects), $\ae=1$ and $\ae=2$.

\newpage

\begin{figure}
\vskip 3. cm
    \includegraphics{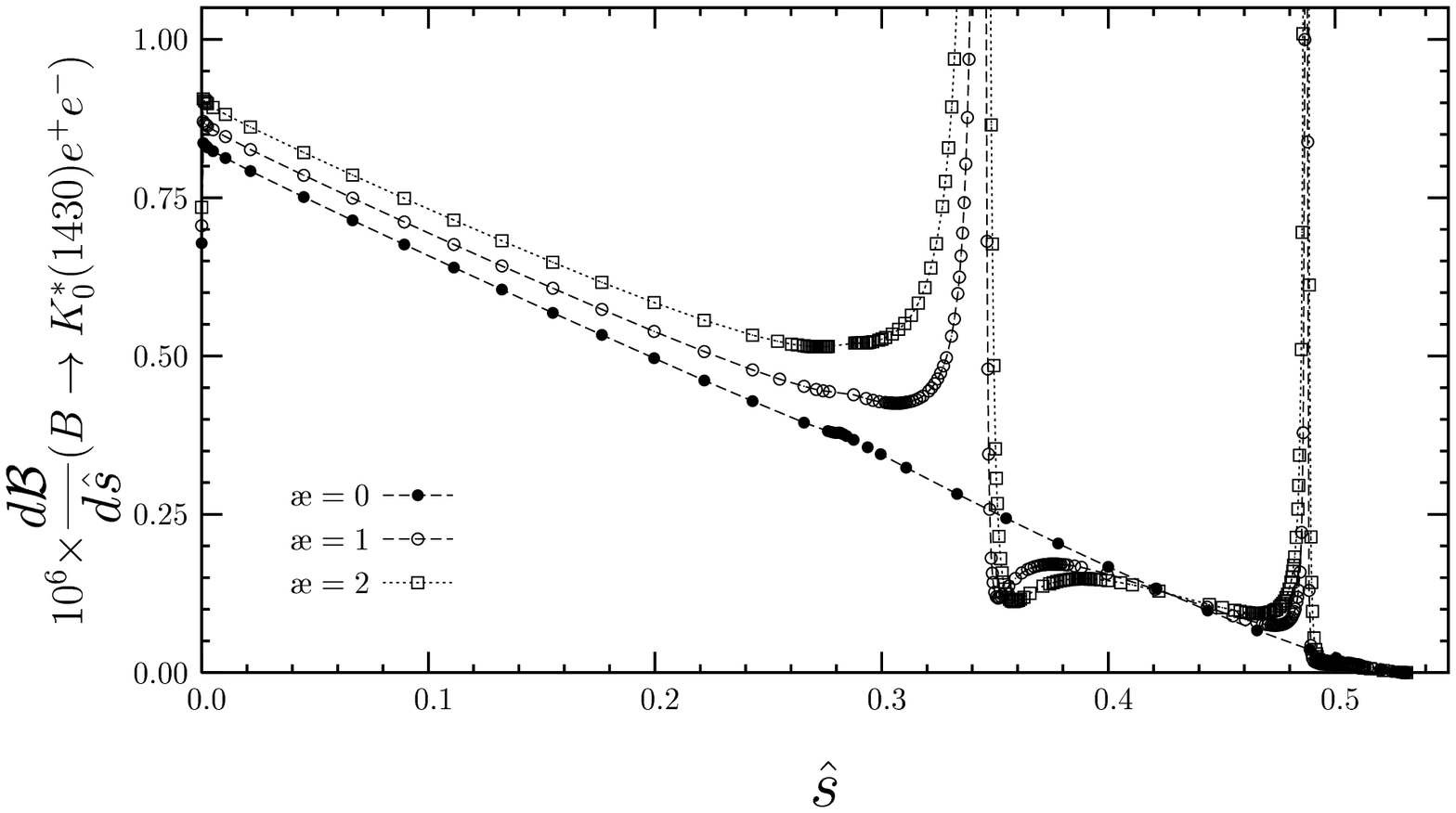}
\vskip 6.3cm
\caption{}
\end{figure}

\end{document}